\documentclass{aa} 
\usepackage{graphicx}
\begin{document}
\title{Spectral differences between the radio-loud and radio-quiet low-hard 
   states of GRS~1915+105: possible detection of synchrotron radiation in 
   X-rays}
\author{S. V. Vadawale\inst{1}, A. R. Rao\inst{1} 
   \and S. K. Chakrabarti\inst{2}}
\offprints{S.V. Vadawale, \email{santoshv@tifr.res.in}}
\institute {Tata Institute of Fundamental Research, Homi Bhabha Road, 
   Mumbai 400 005, India 
   \and S.N. Bose National Center for Basic Sciences, Salt Lake, 
   Calcutta 700091, India}
\date{Received 22 December 2000 /  Accepted 28 March 2001}
\authorrunning{Vadawale et al.}
\titlerunning{Radio-loud and radio-quiet low-hard states of GRS~1915+105}

\abstract{
The Galactic microquasar GRS~1915+105 exhibits several episodes of steady 
X-ray emission characterized by a hard  power-law spectrum  and intense 
Quasi Periodic Oscillations. It is known that there are two types of such 
low-hard states, one with steady radio emission and the other without any 
significant radio emission. We present the results of a detailed X-ray 
spectroscopic study of GRS~1915+105, using data from the {\it Rossi X-ray 
Timing Explorer} obtained during various episodes of the low-hard states 
of the source. We show that there are distinct X-ray spectral differences 
between the radio-quiet and radio-loud low-hard states of the source. The 
X-ray spectra of the radio-quiet low-hard state is best described by a model 
consisting of a multicolor disk-blackbody and a Comptonized component, 
whereas the X-ray spectra of radio-loud low-hard state requires a model 
consisting of three components: a multicolor disk-blackbody, a Comptonized 
component and a power-law, for statistically and physically acceptable fits. 
We attempt to model the presence of this additional power-law component as 
due to synchrotron radiation which is responsible for the radio and infrared 
radiation from the source. We show that a  simple adiabatically expanding jet 
model for the synchrotron radiation can account for the observed X-ray flux 
for reasonable values of the magnetic field and the mass outflow rate. This is 
the first report of detection of the synchrotron radiation in the X-ray band 
for this source.
}

\maketitle

\section{Introduction}

The X-ray transient source GRS~1915+105 was discovered in 1992 (Castro-Tirado, 
Brandt \& Lund \cite{cast:92}) and it earned the name microquasar due to the 
detection of radio lobes moving at apparent superluminal velocities from it 
(see Mirabel \& Rodriguez \cite{mira:99} and references therein). It has been 
in an X-ray bright state ever since its discovery and it shows various types 
of X-ray variability characteristics (Chen, Swank \& Tamm \cite{chen:97}; 
Morgan, Remillard \& Greiner \cite{morg:97}; Muno, Morgan \& Remillard 
\cite{muno:99}; Yadav et al. \cite{yadav:99}; Belloni et al. \cite{bell:00}; 
Rao, Yadav \& Paul \cite{ryp:00}).

Belloni et al. (\cite{bell:00}) found that all the X-ray variability 
characteristics of GRS~1915+105 can be understood as transition from three 
basic states: state A and B characterized by a soft spectrum and state C 
characterized by a hard power-law and intense 0.5 -- 10.0 Hz QPO. The source 
can remain in the state C for long durations, which are defined as class 
$\chi$ (further subdivided into $\chi_1$,  $\chi_2$,  $\chi_3$ and  $\chi_4$) 
by Belloni et al. (\cite{bell:00}). Trudolyubov, Churazov \& Gilfanov 
(\cite{trud:99}) studied the 1996/1997 low-hard state (class $\chi_2$) and 
the state transitions and concluded that the QPO centroid frequency is 
correlated with the spectral and timing parameters and these properties are 
similar to other Galactic black hole candidates in the intermediate state. 
Rao, Yadav \& Paul (\cite{ryp:00}) quantified the spectral states and found 
that state transitions can occur in a very short time in GRS~1915+105.

The $\chi_1$ and $\chi_3$ classes have similar properties characterized by 
high X-ray flux ($\sim$500 mCrab) and high radio flux  ($\sim$ 50 mJy at 2.2 
GHz), which can be associated with the radio-loud ``plateau'' state of the 
source (Fender et al. \cite{fend:99}). The $\chi_2$ and $\chi_4$ classes, on 
the other hand,  have low  X-ray flux ($\sim$250 mCrab) and low  radio flux 
($\sim$ 20 mJy). Recently, Rao et al. (\cite{rao:00}) identified three X-ray 
spectral components during a $\chi_3$ state of the source and found that one 
of the components (the thermal Compton component) is responsible for the QPOs.  
Here we present a detailed X-ray spectroscopic study of the source during 
the steady low-hard states (the $\chi$ class) of the source using data from 
the RXTE archives. We divide the class $\chi$ into two generic sub-classes 
$\chi_{RL}$ (radio-loud) and $\chi_{RQ}$ (radio-quiet) according to the 
observed accompanying radio emission (Fender et al. \cite{fend:99}). The 
sub-class $\chi_{RL}$ includes $\chi_1$ and $\chi_3$ and the sub-class 
$\chi_{RQ}$ includes $\chi_2$ and $\chi_4$ classes. We find a distinct X-ray 
spectral difference between the radio-loud and radio-quiet states of the source
and we attempt to identify one of the X-ray spectral components with the 
synchrotron emission from the source. If confirmed, this will be the first 
detection of X-ray synchrotron emission from this source.

\begin{figure}
   \resizebox{\hsize}{!}{\includegraphics[angle=-90]{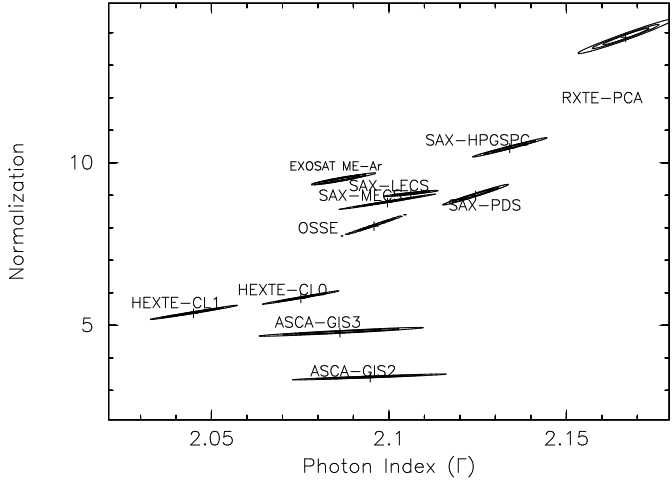}}
   \caption{
    Confidence contours for the normalization and the photon-index for the
    Crab Nebula (and pulsar) in the 0.5 -- 180 keV range obtained from various
    X-ray missions.}
   \label {fig1}
\end{figure}
    
\section{Analysis and Results }

\subsection{Characterization of Instruments}
 {\it Rossi X-ray Timing Explorer (RXTE)} was launched by NASA on 1995 
December 31, with the main objective of timing studies of celestial X-ray 
sources. It has made  great contributions to our understanding of High Energy 
Astrophysics by means of its unrivaled timing resolution. However, here we 
attempt to utilize the excellent opportunity of  wide band hard X-ray 
spectroscopy, provided by the two narrow field of view instruments on board 
the RXTE: PCA (Jahoda et al. \cite{jaho:96}) and HEXTE (Rothschild et al. 
\cite{roth:98}). In order to understand the systematics of the two instruments 
from the end user's perspective, we first analyzed the RXTE archival data of 
the standard candle source, the Crab. We have also analyzed the time averaged 
Crab data (pulsar and nebula taken together) from other missions like BeppoSAX,
ASCA and EXOSAT. Results of our multi-mission study are shown in Figure~
\ref{fig1} (For a detailed discussion on these results see Rao \& Vadawale 
\cite{rv:00}). We find that the wide band (0.5 -- 180 keV) Crab spectra can 
be described by a single power-law with a photon index of 2.1, for most of the
instruments. However, two instruments, HEXTE Cluster 1 and PCA, show systematic
deviation from this value. HEXTE Cluster 1 gives a low photon index. It needs 
some additional systematic errors to give an  acceptable fit and also one of 
the four detectors in this cluster is not capable of giving spectral data. 
Therefore we use data only from HEXTE Cluster 0 in our analysis.

\begin{figure}
   \resizebox{\hsize}{!}{\includegraphics[angle=-90]{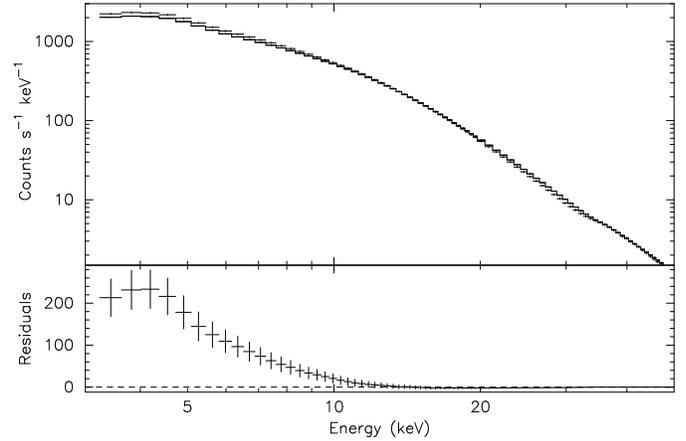}}
   \caption{
   A single power-law fit to the Crab data from RXTE-PCA with photon index 
   fixed at value 2.1. The huge and systematic residuals below 10 keV shows 
   that there is some detector feature which is not properly understood.}
   \label {fig2}
\end{figure}
   
\begin{table*}[t]
\caption{Spectral fits to the Crab data from individual PCUs from RXTE-PCA
with  a  model consisting of a power-law, exponential modification to
the power-law at lower energies and absorption by interstellar medium$^1$.}
\begin{center}
\begin{tabular}{lcccc}
\hline
\hline
Detector & $N_{pow}$ & $A$ & $f$ & $\chi^2$ (dof) \\
\hline
\hline
PCU 0 & $9.27\pm0.12$ & $0.331\pm0.035$ & $0.114\pm0.020$ & 63.33 (81) \\
PCU 1 & $9.53\pm0.12$ & $0.288\pm0.032$ & $0.109\pm0.021$ & 83.34 (83) \\
PCU 2 & $9.79\pm0.13$ & $0.383\pm0.030$ & $0.099\pm0.015$ & 97.35 (81) \\
PCU 3 & $10.50\pm0.15$ & $0.308\pm0.026$ & $0.089\pm0.016$ & 102.50 (84) \\
PCU 4 & $10.45\pm0.13$ & $0.320\pm0.036$ & $0.125\pm0.021$ & 77.23 (78) \\
\hline
\hline
\multicolumn{5}{l}{$^1$$N_H$ is frozen at $0.4\times10^{22}$ cm$^{-2}$ which is 
the average value obtained  from  a  fit}\\
\multicolumn{5}{l}{~~to data from low energy detectors of ROSAT, BeppoSAX and ASCA.}\\
\multicolumn{5}{l}{~~Values of the photon-index and start energy for the
exponential modification} \\
\multicolumn{5}{l}{~~are frozen at 2.1 and 0.5 keV respectively.} \\
\end{tabular}
\end{center}
\end{table*}

The PCA on the other hand gives a high photon index. The problem with PCA is 
a little more than just higher photon index. Figure~\ref{fig2} shows the 
residual for a fit to the Crab data with the photon index frozen at 2.1. 
The existence of a huge excess below 10 keV shows that there might be some 
systematic feature in the instrument which is not properly understood. Trying 
to fit the  data with a single power-law gives an unacceptable fit 
($\chi^2_\nu\approx50$) with a higher photon index and large ``S'' shape 
residuals below 10 keV. The spectral fit can be made acceptable by adding 
a 2 \% systematic error, however the photon index is higher (this was also
noted by Gierlinski et al. \cite{gier:99}) and some residuals below 10 keV 
still appears. Because of this soft excess, any feature below 10 keV should 
be interpreted with caution. Therefore, since we wish to attempt wide band 
spectroscopy, in general we avoid adding components like line or edge which 
have only a local effect and requires high spectral resolution, in the 
spectral analysis with PCA. We find that an exponential modification of the 
form $N(E)=1+Ae^{-fE}$ (where $A$ is amplitude and $f$ is a constant and $E$ 
is energy in keV -- model {\it expfac} in XSPEC), at low energies to the 
power-law with photon index 2.1 can explain the Crab data from PCA. Parameters 
of this model can be used to quantify the soft excess. We have checked whether 
this soft excess can be attributed to any particular Proportional Counter 
Unit (PCU). Table 1 shows that this soft excess, of the same order, is present
in all the PCUs and it is not possible to identify a few of them as good 
detectors. We have also verified that the results obtained from wide-band 
spectral fitting does not depend on selection of a combination of PCUs.
Therefore in our wide band spectral analysis we use data from all PCUs added 
together with 2\% systematic error along with the data from HEXTE Cluster-0.
	 
\subsection{Data reduction and Spectral fitting}
	We have selected representative observations for each sub-class
$\chi_{RQ}$ and $\chi_{RL}$. Table 2 gives the details of these individual
observations. The data reduction and analysis was performed with the software
FTOOLS (Ver 5) and XSPEC (Ver 11). We extracted 129 channel spectra from PCA 
Standard-2 data (data from all PCUs added together) and 64 channel spectra
from HEXTE Cluster-0 archive data. We generated the corresponding background
spectra and response matrices following the standard procedures. We fitted 
the PCA (3 -- 50 keV) and HEXTE-CL0 (15 -- 180 keV) spectra simultaneously 
with different spectral models such as disk-blackbody; disk-blackbody and 
comptonization due to hot plasma (CompST - see Sunyaev \& Titarchuck 
\cite{sun:80}); disk-blackbody and cutoff power-law; disk-blackbody, CompST 
and power-law; etc. modified by interstellar absorption. We fixed the value 
of the interstellar absorption to $N_H=6\times10^{22}$ cm$^{-2}$ (Belloni 
et al. \cite{bell:97}; Markwardt et al. \cite{markw:99}; Muno et al. 
\cite{muno:99}) and have used the normalization of HEXTE spectrum (with 
respect to the PCA spectrum) as a free parameter so as to account for any 
residual uncertainties in the relative area of the two instruments. For each 
observation, the values of $\chi^2$ (degrees of freedom -- dof) for different 
models are given in Table 2A. The spectral parameters for the best fit model 
are given in Table 2B.

 We find that addition of a Gaussian feature for iron K$_\alpha$ line near 
6.4 keV in our models does reduce the $\chi^2$ values but does not have any 
effect on the spectral parameters of the other components. For example, for 
the data obtained on 1997 October 16, addition of a line feature reduces the 
value of $\chi^2_\nu$ to 1.2 (from 1.6), but the values of the parameters 
remain almost the same (within a few percent of the values). The derived value 
of the equivalent width is rather high ($\sim$150 eV). The high spectral 
resolution observation of this source using ASCA or Chandra show very low 
equivalent width (10 -- 20 eV) for iron K$_\alpha$ line (Kotani et al. 
\cite{kota:00}; Lee et al. \cite{lee:01}). This suggests that the observed 
features might be mainly due detector artifact.  Further, the addition of line 
feature affects all the continuum models in almost identical way with similar 
reduction in the value of $\chi^2$. Since it does not affect our inference 
based on the wide band structure of the spectrum, we do not include the iron 
line in our fits. 

\begin{figure}
   \resizebox{\hsize}{!}{\includegraphics{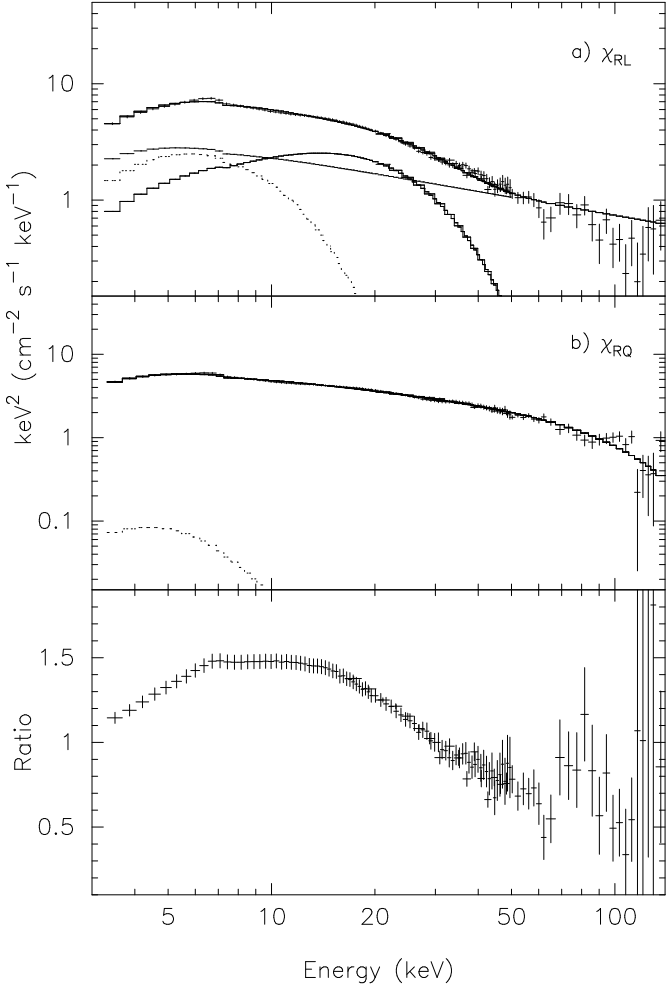}}
   \caption{ 
    The deconvolved spectra during the $\chi_{RQ}$ and $\chi_{RL}$ states of 
    GRS~1915+105 are shown in the top two panels of the figure. The model 
    consists of a disk blackbody (shown as dotted lines) and emission from 
    a Compton cloud. The $\chi_{RL}$ state requires an additional power-law 
    component. The bottom panel shows the ratio of observed count rates 
    ($\chi_{RL}$ to $\chi_{RQ}$ ratio), which highlights the spectral 
    difference in the 15 -- 50 keV range between the two states.}
   \label {fig3}
\end{figure}

\subsection{Radio-quiet low-hard state}
	The standard model for the Black Hole Candidates (BHCs) consists 
of a multicolour disk-blackbody and a power-law. The power-law is generally
assumed to be an approximation for some physical processes like Comptonisation
of soft photons from a hot plasma. Since we have used wide band data upto 
180 keV, we find that a power-law does not give a satisfactory fit to the 
X-ray spectra of the radio quiet low-hard state (Table 2), particularly because
there are fewer counts at high energies ( $>$ 70keV) than are predicted by 
the model. A model consisting of a disk-blackbody and a cutoff power-law gives 
a better fit, however, the best fit is obtained for a model consisting of 
a disk-blackbody and a CompST. The CompST appears to be the best candidate for 
the second component also due to the fact that it describes the actual process 
of the inverse compton scattering of the soft photons. The high energy cutoff 
also comes naturally in the CompST. Therefore we conclude that the 3 -- 180 keV
spectra of the radio-quiet low-hard state is best described by the model 
consisting of a disk-blackbody and a CompST. It should be noted here that 
the wide band spectral fitting is critical to distinguish these models. It 
should also be pointed out here that the CompST model does not include 
relativistic corrections and hence more sophisticated models may be required 
if we include data above 200 keV.

\subsection{Radio-loud low-hard state}
	The X-ray spectra of the radio loud low-hard state is peculiar and
needs a careful interpretation. Any combination of two spectral components
gives a statistically inferior fit to the data. On examination of the residuals
it is found that this is mainly due to the fact that the data has a spectral 
curvature in the 15 -- 40 keV region and a flattening above 70 keV, thus 
necessitating at least two spectral components above 15 keV (or three component
in the 3 -- 180 keV region). The improvement in the value of $\chi^2$ is more 
than 100, thus conclusively showing that a combination at least three 
components are required to adequately explain the wide band data of the 
radio-loud low-hard state of GRS~1915+105. 
	This shows that the wide band X-ray spectrum of the radio-loud low-hard
state is qualitatively different from that of the radio-quiet low-hard state. 
To highlight this difference we have shown in Figure~\ref{fig3} the unfolded 
spectrum of the radio-loud low-hard state obtained on 1997 October 16 (top 
panel), radio-quiet low-hard state obtained on 1996 December 4 (middle panel) 
and the ratio of the observed count rates in the radio-loud state to those in 
the radio-quiet state in each energy channel (normalized to the value at 30 
keV), in the bottom panel. The spectral curvature in the 15 -- 45 keV region 
can be seen even in the raw data. 

 	In the fit with a disk-blackbody and a power-law, this spectral 
curvature in the $\chi_{RL}$ class is mimicked by the disk-blackbody, giving an 
unrealistically high temperature (4 -- 5 keV). Further, the resultant values of
$\chi^2$ are inferior to that obtained for the 3 component fit. Fit with only 
disk-blackbody  and CompST is also not acceptable, now because CompST decreases
much more sharply and there are more counts at higher energies ($>$70 keV) than
predicted. Thus an additional power-law is needed to account for them. It 
should be noted here that, though the existance of the three components is 
fairly certain here, the exact values of parameters of disk-blackbody and 
CompST should be used with caution, because of the large systematic error 
added to the PCA data.

 The wide band X-ray spectra of GRS~1915+105 in the radio-loud state has been 
reported earlier by Muno et al. (\cite{muno:99}). This is the only reported 
X-ray spectra of radio-loud low-hard state of GRS~1915+105. They use a model 
consisting of disk blackbody, power-law and a gaussian line corresponding
to iron K$_\alpha$. We get similar spectral parameters when we use disk 
blackbody and power-law as the continuum models. As mentioned earlier, we get 
a better fit using the three component model. Further, the line equivalent 
width is quite high ($\sim$200 eV) compared to that observed with ASCA or 
Chandra. The derived parameters of the disk-blackbody (inner disk temperature 
$kT_{in} \sim$ 4 -- 5 keV and normalization $N_{bb} \sim$ 0.5 - 0.7) appear 
to be quite unrealistic.  Such a high inner disk temperature is not observed 
in any other X-ray binary. The value of $N_{bb}=0.7$ corresponds to an inner
disk radius $R_{in}\approx1.8$ km for this source 
($R_{in,km} = D_{10kpc}\sqrt{N_{bb}/\cos{\theta}}$, D=12.5 kpc and $\theta=70$).
The inner disk radius of 1.8 km is clearly unphysical because the Schwarzschild
radius for a one solar mass black hole is $\sim$3 km. 

\begin{table*}[t]
\caption{Results of spectral analysis of RXTE data of GRS~1915+105 during 
radio-loud and radio-quiet low-hard states}
\begin{center}
\scriptsize
\begin{tabular}{lccccccc}
\multicolumn{7}{l}{{\bf A.} Observation log, radio flux$^1$ and $\chi^2$ 
(dof) for different models$^2$} \\
\hline
\hline
Date & PID &~~~Class~~~& Radio flux (mJy) &\multicolumn{4}{c}{$\chi^2$ (dof)}\\
 & & & (at 2.25 GHz) & dbb+po & dbb+ctpo & dbb+co & dbb+co+po \\
\hline
\hline
1997 Oct 16 & 20402-01-50-01 & $\chi_{RL}$ & 51.5 & 404.23(124) & 
259.31(123) & 304.59(123) & 188.03(121) \\
1998 May 12 & 30402-01-12-01 & $\chi_{RL}$ & 68.6 & 366.46(124) & 
233.94(123) & 276.21(123) & 171.42(121) \\
1999 Jun 7 & 40703-01-17-01 & $\chi_{RL}$ & 31.4 & 221.85(115) & 
176.87(114) & 249.45(114) & 140.42(112) \\
1996 Dec 4 & 20402-01-05-00 & $\chi_{RQ}$ & 13.8 & 189.26(124) & 
149.84(121) & 137.41(121) &  --  \\
1998 Aug 31 & 30703-01-31-00 & $\chi_{RQ}$ & 13.4 & 151.61(117) & 
106.56(116) & 99.18(116) &  --  \\
\hline
\hline
\\
\multicolumn{7}{l}{{\bf B.} Best fit X-ray spectral parameters$^3$ and 
2-50 keV X-ray flux of individual spectral components } \\
\hline
\hline
Date & $kT_{in}$ & $kT_e$ & $\tau$ & $\Gamma_x$ &\multicolumn{3}{c}{2-50 keV
Flux (10$^{-9}$ erg~cm$^{-2}$~s$^{-1}$) }\\
 & (keV) & (keV) & & & Disk Blackbody & CompST & Power-law \\
\hline
\hline
1997 Oct 16 & 1.99$^{+0.05}_{-0.04}$ & 4.89$^{+0.09}_{-0.08}$ & 
10.25$^{+0.30}_{-0.27}$ & 2.49$^{+0.01}_{-0.01}$ & 4.65 & 7.78 & 9.41 \\
1998 May 12 & 1.73$^{+0.03}_{-0.03}$ & 4.08$^{+0.08}_{-0.08}$ & 
12.60$^{+0.47}_{-0.59}$ & 2.61$^{+0.04}_{-0.04}$ & 7.93 & 5.50 & 11.33 \\
1999 Jun 7 & 1.52$^{+0.10}_{-0.10}$ & 6.73$^{+0.23}_{-0.21}$ & 
5.82$^{+0.16}_{-0.16}$ & 2.59$^{+0.02}_{-0.02}$ & 1.96 & 12.87 & 9.33 \\
1996 Dec 4 & 1.28$^{+0.29}_{-0.36}$ & 20.76$^{+1.11}_{-0.97}$ & 
2.94$^{+0.10}_{-0.10}$ & -- & 0.34 & 20.53 & -- \\
1998 Aug 31 & 1.13$^{+0.09}_{-0.08}$ & 21.96$^{+3.11}_{-2.19}$ & 
2.78$^{+0.21}_{-0.24}$ & -- & 2.23 & 19.13 & -- \\
\hline
\hline
\multicolumn{8}{l}{$^1$Average flux for the respective day, from Green Bank 
Interferometer public data} \\
\multicolumn{8}{l}{$^2$The model components are  {\bf dbb:} disk-blackbody, 
{\bf po:} power-law, {\bf ctpo:} cutoff power-law, {\bf co:} CompST } \\
\multicolumn{8}{l}{$^3$The spectral parameters are: 
~~$kT_{in}$: Inner disk temperature  
~~$kT_e$: Electron temperature of the Compton cloud}\\
\multicolumn{8}{l}{~~~~~~~~~~~~~~~~~~~~~~~~~~~~~~~~~~~~~~~~~~~
$\tau$: Optical depth of the Compton cloud
~~$\Gamma_x$: Photon index} \\
\end{tabular}
\end{center}
\end{table*}

\section{Origin of the  additional Power-law}
 
 The previous section gives a consistent picture according to which, the 
disk-blackbody and CompST components are always present in the X-ray spectra
of the low-hard state. During the low-hard state accompanied by high radio 
emission, the X-ray spectra requires an additional power-law component. The 
2 -- 50 keV X-ray flux of the additional power-law also shows some correlation 
with the accompanying radio flux (see Table 2B). This naturally leads to 
a hypothesis that the additional power-law component is also due to the same 
mechanism i.e. synchrotron radiation, which is responsible for the observed 
radio emission. Dhawan et al. (\cite{dhaw:00}) have reported an AU scale radio 
jet from the core of the source during the $\chi_3$ state of the source 
observed in 1997 October. Recently Markoff et al. (\cite{markf:01}) showed 
that the entire X-ray emission from a black hole binary XTE~J1118+480 can be 
explained as synchrotron emission from the innermost part of a jet. Hence it 
is quite conceivable that the additional power-law component seen in X-rays 
is emitted from the base of such a jet. 

\subsection{A synchrotron jet model}
Here we present a simplified model similar to that given by Hjellming \& 
Johnston (\cite{hjel:88}), to examine whether the power-law component can 
indeed be explained as a synchrotron emission from the base of a compact, 
adiabatically expanding conical jet in GRS~1915+105. Since we are interested 
in the synchrotron radiation in X-rays, we consider emission from the entire 
volume of the jet instead of emission from only the outer layer of the jet as 
done by Hjellming \& Johnston (\cite{hjel:88}), who were mainly interested in 
the radio emission. 

We make the following assumptions: 

\begin{enumerate}
\item The mass outflow from regions close to the black-hole is conical in 
shape with a small opening angle $\theta$ ($\sim~10^{\circ}$).
\item Electron-proton plasma is injected at the base of the jet with 
relativistic velocities.
\item Electron acceleration occurs at the shock front located at distance
$z_{s}$ from the base, resulting in the power-law distribution in energy
$N_e=kE^{-p}$.
\item There is no mass loss across the sides of the jet.
Magnetic field and particle distribution are uniform across any 
cross-section of the jet.
\end{enumerate}

 Let $B_s$ and $k_s$ be the magnetic field and proportionality constant of 
the power-law electron distribution at the shock front, and $r_s$ be the 
radius of the cone at the shock front given by 
$r_s=r_0+z_{s}\tan(\theta/2)$, where $r_0$ is the base radius of the jet
assumed to be 100 Schwarzschild radius. Then from conservation of magnetic 
flux along the jet, we get (see Hjellming \& Johnston \cite{hjel:88})

\begin{center}
\begin{equation}
\label{Bl}
B_{\ell}=B_s\left( \frac{r_{\ell}}{r_s}\right) ^{-1} 
\end{equation}
\end{center}

 where $r_{\ell}$ is the radius of the cone at a distance $\ell$ from the base,
given by

\begin{center}
\begin{equation}
\label{Rl}
r_{\ell}~=~r_0~+~\ell\tan(\theta/2)
\end{equation}
\end{center}

Further, if we assume that the expansion with $r_{\ell}$ is adiabatic, then
conservation of both the total energy and the number of relativistic particles
requires that  (see Hjellming \& Johnston \cite{hjel:88})

\begin{center}
\begin{equation}
\label{Kl}
k_{\ell}=k_s\left( \frac{r_{\ell}}{r_s}\right) ^{-2(p+2)/3}
\end{equation}
\end{center}

 The values of $B_{\ell}$ and $k_{\ell}$ determine the volume emissivity 
$J_{\ell}(\nu)$ and absorption coefficient $\chi_{\ell}(\nu)$ of the local 
medium (Longair \cite{long:94})

\begin{eqnarray}
\label{Jl}
J_{\ell}(\nu) & = & 2.344\times10^{-25} a(p) k_{\ell} B_{\ell}^{(p+1)/2} 
\nonumber\\ 
 & & \times \left(\frac{1.253\times10^{37}}{\nu} \right)^{(p-1)/2}  ~~~~~~
 \mathrm{W m^{-3} Hz^{-1}}
\end{eqnarray}

\begin{eqnarray}
\label{Chil}
\chi_{\ell}(\nu) & = & 3.354\times10^{-9} b(p) k_{\ell} B_{\ell}^{(p+2)/2} 
\nonumber \\
 & & \times (3.54\times10^{18})^p \nu^{-(p+4)/2}  ~~~~~~~~~~~~~\mathrm{m^{-1}}
\end{eqnarray}

 where $a(p)$ and $b(p)$, the functions of the the particle 
power-law index $p$ involving numerous $\Gamma$ functions, are of the order of 
unity. We consider the $\chi_{RL}$ state observed during 1997 October, for
which spectral index is $\sim$ 1.5 (photon index $\sim2.5$) which gives $p=4$.
The values of $a(p)$ and $b(p)$, for $p=4$ are 1.186 and 0.230 (Longair 
\cite{long:94}, Table 18.2).

 Consider a segment of the cone with length $d\ell$ at the distance $\ell$ 
from the base. This segment can be approximated by a cylinder of length 
$d\ell$ and radius $r_{\ell}$. Now for this cylinder, consider a shell of 
thickness $dr$ at the radius $r$ from the center.  Total self absorbed 
synchrotron emission from this shell is (see Longair \cite{long:94})

\begin{equation}
\label{Is}
I_s(\nu) = 2 \pi r d\ell \frac{J_{\ell}(\nu)}{4\pi\chi_{\ell}(\nu)}(1-e^{-\chi_{\ell}(\nu)dr})
\end{equation}

where $2 \pi r d\ell$ is the area of the outer surface of the cylinder.
This synchrotron radiation is absorbed in the medium of thickness 
$(r_{\ell} - r)$. Hence the observed spectrum of the radiation from this 
shell is 

\begin{equation}
\label{Ir}
I_r(\nu)~=~I_s(\nu)e^{-\chi_{\ell}(\nu)(r_{\ell}-r)}
\end{equation}

 Total observed spectrum from this segment is

\begin{equation}
\label{Il}
I_{\ell}(\nu)~=~\int_{0}^{r_{\ell}}I_r(\nu)dr
\end{equation}

 and the total emission spectrum from the entire conical jet is

\begin{equation}
\label{Iv}
I(\nu)~=~\int_{0}^{L}I_{\ell}(\nu)d\ell
\end{equation}

 where $L$ is the assumed length of the jet. This is in the rest frame of the
jet. To get the observed flux in the rest frame of the observer we have to 
transform it according to the transformations (Falcke \& Biermann 
\cite{fal:96}),

\begin{equation}
\nu_{obs} = D\nu, ~~~~~~~~I_{obs}(\nu_{obs})= D^2 I(\nu_{obs}/D) 
\end{equation}

Where $D=[\gamma(1-\beta\cos\Theta)]^{-1}$ is the Doppler factor of the jet
matter moving at the relativistic velocity $\beta c$ at angle $\Theta$ from 
the line of site. We assume them to be 0.9 c and 70$^{\circ}$ respectively.

 Thus the observed flux is

\begin{equation}
\label{Fv}
F(\nu)~=~\frac{D^2 I(\nu) \sin\Theta}{4\pi d_{sc}^2}
\end{equation}

 where $d_{sc}$ is the distance to the source, which we assume to be 12 kpc 
(Mirabel \& Rodriguez \cite{mira:94}).

We take  a  numerical approach to solve this radiation transfer problem. We 
divide the entire jet length in large a number of cylindrical segments 
($\sim$100000), calculate the synchrotron emission from each segment assuming 
uniform $B_{\ell}$ and $k_{\ell}$ for each segment and finally get the observed
flux by integrating over the jet length. The basic free parameters of this 
model are the magnetic field $B_s$, the proportionality constant $k_s$ at the 
shock front, the distance of the shock front from the base $z_{s}$. The value 
$k_s$ in itself is not physically meaningful, however it can be converted 
into a meaningful quantity like mass outflow rate. The total electron number 
density of the accelerated electrons, $N_e$ is given by 
$k_{\ell}\int_{E_{min}}^{E_{max}}E^{-p}dE$. For $p>2$ it is governed by 
$E_{min}$. Assuming that 50 \% of the electrons are accelerated at relativistic
energies and that there is one proton per electron, mass density is given by 
$\rho=2 N_e \gamma m_p$. Then the total mass outflow rate is 
$\dot{M}_o=\beta c \rho\pi^2R_s$ (Falcke \& Biermann \cite{fal:96}). Thus, 
in order to get mass outflow rate $\dot{M}_o$, the most important assumptions 
we have to make are about the lower energy cut-off of the electron energy 
distribution and velocity of the jet. 

\begin{figure}
   \resizebox{\hsize}{!}{\includegraphics[angle=-90]{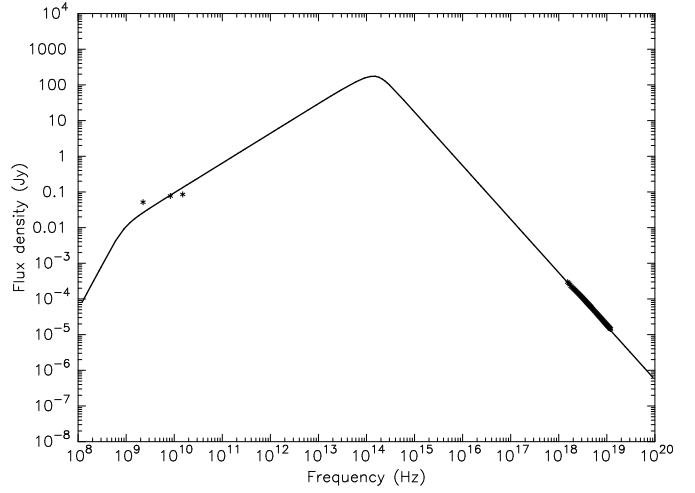}}
   \caption{
   A simplified model spectrum of synchrotron emission from a conical jet 
   of length 100 AU with a magnetic field of $5\times10^3$ G at the shock 
   front. The stars represent observed fluxes on 1997 October 16, at 2.25 GHz, 
   8.3 GHz (GBI public data), 15 GHz (Fender et al. 1999) and 20 -- 150 keV 
   (present work).}
   \label{fig4}
\end{figure}

\subsection{Parameter space}
The predicted emission spectrum from the jet of length 100 AU is shown in 
Figure~\ref{fig4} (solid line) along with the observed points in X-ray as well 
as radio during the $\chi_{RL}$ state of 1997 October. The parameters used in 
this calculation are $B_s=5\times10^3$ G, $k_s=1.2\times10^{-13}$, 
$z_{s}=10^5~R_{sw}$ (Schwarzschild  radius) and $\theta$ = 10$^{\circ}$. 
This value of  $k_s$ gives mass density $\rho_s$ at the shock to be 
$\sim3.7\times10^{-11}$~kg~m$^{-3}$, assuming lower energy cutoff 
$E_{min}=1 GeV$ ($\gamma_{min}\approx2000$) which requires the mass outflow 
rate of $\dot{M}_o=2.3\times10^{15}$~kg~s$^{-1}$ for outflow velocity of 0.9 c.
Total jet power given by these parameters is $4.28\times10^{39}$~erg~s$^{-1}$. 
Both the magnetic field $B_s$ and $\dot{M}_o$ are within the possible range 
(see next section). Taking $\gamma_{min}\approx100$  increases the required 
mass outflow rate by three order of magnitude.

As can be seen from equation \ref{Jl}, the  same flux at any frequency can be 
obtained for various values of magnetic fields and electron densities i.e. 
mass outflow rates. The possible parameter space to get the observed X-ray 
flux is shown in Figure~\ref{fig5}. In this calculation we have assumed the 
values of many parameters, like velocity of the jet, jet angle, ratio of number
density of the relativistic electrons to the total electron number density, 
base radius, distance of the source and angle between jet and line-of-sight. 
Among these, the parameters on which the mass outflow rate depends most 
sensitively are the velocity of the jet, ratio of number density of the 
relativistic electrons to the total electron number density, and the base 
radius. The values of distance of the source and the inclination angle are 
well known from the literature. Also, the final result does not depend on 
these parameters very sensitively. Among the remaining three, the assumption 
about the jet velocity is on the safer side, however, for the other two 
parameters we have assumed values which seem intuitively reasonable.
It should be pointed out here that not all shown combinations of magnetic 
field and mass outflow rates produce radio fluxes similar to the observed flux 
because the turnover frequency is slightly different for different combinations.
Also the turnover frequency strongly depends on the electron energy index, $p$.
The observed radio data suggests a turnover frequency in the range 
$10^{14}-10^{15}$ Hz. For physically reasonable magnetic field and mass outflow
rates, this turnover frequency range can be obtained only for a narrow range 
of electron energy index around 4.0. Thus, apart from the observed X-ray 
power-law, the observed radio flux also requires a high electron energy index.

\section{Discussion}

 Synchrotron emission from a conical jet was first studied by Blandford \& 
Konigl (\cite{bk:79}) in the context of AGNs. Reynolds (\cite{rey:82}) explored
in more detail the observed spectra form winds and jets with a variety of 
geometries, magnetic fields and energetics. Hjellming \& Johnston 
(\cite{hjel:88}) discussed the application of such models to X-ray binaries. 
Extending this further, Falcke \& Biermann (\cite{fal:96}; \cite{fal:99}) 
developed  a disk-jet 'symbiosis' model to explain the radio emission from 
various accretion powered sources, according to which some fixed fraction of 
the accreted mass comes out as a jet. However, these models are developed 
mostly to account for the observed radio emission and does not discuss 
synchrotron emission in the high frequency range. Recently, Markoff et al. 
(\cite{markf:01}) showed that, in a black hole binary, synchrotron emission 
can give observable X-ray flux. They showed that the entire X-ray flux of the 
black hole binary XTE~J1118+480 can be explained as synchrotron emission. For 
XTE~J1118+480 almost all the emission from the accretion disk is in the EUV 
range. In GRS~1915+105, however, the X-ray flux has significant contributions 
from the accretion disk as well as the inner Compton could. We have separated 
the flux from the additional power-law component and have attempted to identify
it with the synchrotron emission and correlate with the observed radio emission. 

\begin{figure}
   \resizebox{\hsize}{!}{\includegraphics[angle=-90]{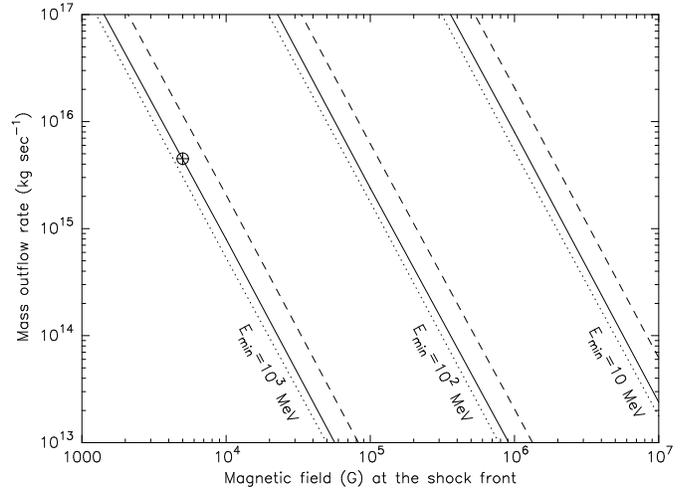}}
   \caption{
   Mass outflow rate vs. Magnetic field at the shock front required to give 
   the observed X-ray flux. The dashed lines represent the distance of the 
   shock front $z_{s}=5\times10^4~R_{SW}$, the solid lines represent 
   $z_{s}=10^5~R_{SW}$ and the dotted lines represent 
   $z_{s}=1.5\times10^5~R_{SW}$. Three groups of this lines are shown for
   different lower energy cutoff to the electron energy distribution. The 
   circle shows the parameters used in the calculation of spectrum 
   of Figure~\ref{fig4}.}
   \label {fig5}
\end{figure}

It can be seen from Figure~\ref{fig4} that the observed X-ray power-law agrees 
very well with that predicted by this simple synchrotron model, though the 
parameter space shown in Figure~\ref{fig5} is a bit on extreme side. For 
the $\chi_{RL}$ state under consideration (1997 October), the inner disk 
temperature ($T_{in}$) of 2 keV gives  a  mass accretion rate  of 
$\dot{M}_{in}\sim10^{16}$~kg~s$^{-1}$. If we assume the outflow rate to be 
$\sim$10~\% of $\dot{M}_{in}$ (such outflows from very close to the compact 
object have been predicted by Chakrabarti (\cite{chakb:99}) and Das \& 
Chakrabarti (\cite{das:99})), then the minimum required magnetic field strength
at the shock front is $\sim10^4$~G. If we extend the same dependence back to 
the base of the jet, then the base magnetic field required is $\sim10^6$~G. 
The base magnetic field value of the order of 10$^6$ G looks a bit higher, 
however, it should be noted that the magnetic field values of the order of 
0.1 G, reported in the literature (Dhawan et al. \cite{dhaw:00}; Fender et al. 
\cite{fend:99}) are at a very large distance from the base and lead to a 
similar value if extrapolated back to the base. Markoff et al. (\cite{markf:01})
have also reported that a magnetic field $\sim10^6$ is required to explain 
the X-ray emission as synchrotron radiation, in the black hole binary 
XTE~J1118+480. The scaling laws given by Narayan et al. (\cite{nar:98}) also 
predict a similar value for the equi-partition magnetic field. This means that 
the base magnetic field values from 10$^5$ to 10$^7$ G and corresponding mass 
outflow can produce the observed X-ray flux. The location of the shock, 
$z_s$, also appears to be large, however, such a distance is required from 
the consideration of electron life-time in high $B$. Figure~\ref{fig5} shows 
that as $z_s$ decreases the required $B_s$ increases for a given $\dot{M}_o$. 
Now a base magnetic field $B_0$ of the order of $10^5 - 10^6$ is expected, as 
discussed above. This makes $B_s$ very large for smaller $z_s$. The energy 
loss rate of electrons of energy $E$ in magnetic field $B$ is given by 
$-(dE/dt) = 1.058\times B^2\gamma^2 (v/c)^2$ W.
For $B\sim10^5$ this gives the life-time of an electron to be less than 
0.1 second for any electron energy. Thus in such a strong magnetic field all 
electrons would thermalise immediately, which is not the case for this source,
for which the radio emission is observed at $\sim$ 10 AU (Dhawan et al. 
\cite{dhaw:00}). This suggests that the electron acceleration site must be far
away from the base where the magnetic field is moderately strong.

Figure~\ref{fig4} also shows that this model can predict the radio flux similar
to that of the observed flux, however, the predicted spectral index in this 
frequency range is steeper than the observed value. According to this model, 
the spectral index in the optically thick region depends on the index of the 
electron energy distribution and is given by $\alpha_{thick}=10(p-1)/(7p+8)$ 
(see Hjellming \& Johnston \cite{hjel:88} for derivation), which for present 
case (p=4), is $\alpha_{thick}=0.833$ whereas the observed spectral index is 
$\sim$0.3 during this observation. There are other models (Falcke  
\cite{falh:96}; Falcke \& Biermann \cite{fal:99}) which predict a much flatter 
spectral index in the optically thick region which and are also independent of 
the electron energy index. However, these problems are only for the optically 
thick frequencies i.e. frequencies below the turnover frequency. For optically 
thin frequencies, this model agrees with the observed data and hence the 
observed additional power-law component in the $\chi_{RL}$ states can indeed 
be due to the synchrotron radiation for a wide range of physically meaningful 
parameters.

 Recently Naik \& Rao (\cite{naik:00}) have shown that, of the 12 classes in 
GRS~1915+105 as defined by Belloni et al. (\cite{bell:00}), high radio emission
is present only during three classes: $\chi$, $\beta$ and $\theta$. The source 
is in the low-hard state during a  significant fraction of time for both the 
classes $\beta$ and $\theta$. We have verified that the spectral properties of 
the low-hard states during these states are also similar to the $\chi_{RL}$ 
state reported here. Thus it appears that, whenever there is a significant 
radio emission from the source (particularly with a flat or an inverted 
spectrum), it manifests itself in the X-ray spectra as an additional power-law
component. 

 One more striking difference between the X-ray spectra of these two states
is the different shape of CompST. The temperature of the Compton cloud,
responsible for the Comptonized component, is lower in the radio-loud state,
whereas the optical depth is much larger. Such a large optical depth has 
been reported earlier for this source (Rao et al. \cite{rao:00}). We have 
verified that optical depth of the order of unity can be rejected for the 
radio-loud state with very high statistical confidence (change in $\chi^2$ of
$>$ 200). We find that the spectral curvature at 15 -- 40 keV needs a large 
optical depth, if this curvature is due to a Comptonization model. It should 
be noted here that according to the TCAF model of Chakrabarti \& Titarchuk 
(\cite{chakb:95}), the optical depth of the post-shock region should be of the 
order of the accretion rate (normalized to the Eddington accretion rate) and 
hence should be less than 1. The observations reported here pertain to the 
case of high radio emission and related to an accretion disc with very high 
mass out-flow rate. It is quite possible that the region responsible for the 
thermal-Compton emission is a geometry where material outflow, Compton cooling
and Comptonization of the disk emission is taking place and the parameters 
derived by the simple thermal-Compton model could be an approximation for the 
emission from such a complex region. Das \& Chakrabarti (\cite{das:99}) have 
proposed a combined inflow/outflow model in which they show that the outflow 
rate depends on the inflow parameters. It would be interesting to investigate 
the relation between the different Compton cloud properties in the presence of
outflow.

\section{Conclusions}

	Here we have shown that the X-ray spectra during the radio-loud and 
radio-quiet low-hard states of the Galactic microquasar GRS~1915+105 are 
different. The most striking difference is the presence of an additional 
power-law component. We have shown that this component can arise due to the 
synchrotron radiation, by comparing the observed X-ray flux to the numerically 
estimated synchrotron spectrum from accelerated electrons at the shock front 
in a simple adiabatically expanding jet.  Considerably wide ranges of mass 
outflow rate and magnetic fields at the shock can produce the observed X-ray 
power-law flux. This model can also predict the observed radio flux, however,
the predicted spectral index of the radio emission is different than the 
observed spectral index.

\begin{acknowledgements}

 We thank G. C. Dewangan, B. Paul, M. Choudhury, and J. S. Yadav for very 
useful discussions.
 This research has made use of data obtained through the High Energy
Astrophysics Science Archive Research Center Online Service, provided by the
NASA/Goddard Space Flight Center. 
 The Green Bank Interferometer is a facility of the National Science 
Foundation operated by the NRAO in support of NASA High Energy Astrophysics 
programs.

\end{acknowledgements}

\subsection*{Note added in proof}
It was pointed out to us that the total radiated power from the jet (by 
integrating the predicted spectra in Figure~\ref{fig4}) comes more than the 
Eddington luminosity of this source ($\sim10^{39}$ erg~s$^{-1}$). Such a high 
radiated power from the jet is unlikely. However, this is because of the rather
steep spectral index in the optically thick region which causes a high flux at
the turnover frequency. The observed flux in the radio band is $\sim10^{31}$ 
erg~s$^{-1}$ and the observed power in X-rays due to the jet (as suggested 
here) is $\sim10^{38}$ erg~s$^{-1}$. Hence the high radiated power in the jet 
is mainly due to the large peak at $\sim10^{14}$ Hz. A more realistic model 
which directly connects radio to X-rays by (a)  incorporating models which 
predict flatter spectral index in the radio band (Falcke \cite{falh:96};
Falcke \& Biermann \cite{fal:99}) and (b) a model incorporating energy loss 
which will have one more spectral turn over at intermediate frequencies, might 
be able to give more realistic value for radiated jet power. It should also 
be noted that Gliozzi et al. (1999; MNRAS 303, L37) have estimated a power in 
kinetic bulk motion of GRS 1915+105, which far exceeds the Eddington luminosity.

{}

\begin{thebibliography}{}

\bibitem[1997]{bell:97}
Belloni, T., Mendez, M., King, A. R., van der Klis, M., \& van Paradijs, J.
1997, ApJ, 488, L109

\bibitem[2000]{bell:00}
Belloni, T., Klein-Wolt, M., Mendez, M., van der Klis, M., \& van Paradijs, J. 
2000, A\&A,  355, 271

\bibitem[1979]{bk:79}
Blandford, R. and Konigl, A. 1979, ApJ, 232 34

\bibitem[1992]{cast:92}
Castro-Tirado, A. J., Brandt, S., \& Lund, N. 1992, IAU Circ., 5590

\bibitem[1999]{chakb:99}
Chakrabarti, S. K. 1999, A\&A, 351, 185

\bibitem[1995]{chakb:95}
Chakrabarti, S. K., \& Titarchuk, L. G. 1995, ApJ, 455, 623.

\bibitem[1997]{chen:97}
Chen, X., Swank, J. H., \& Taam, R. E.  1997, ApJ, 477, L41

\bibitem[1998]{das:99}
Das, T. K. \& Chakrabarti, S. K. 1999,  {\it Classical and Quantum Gravity}, 
V. 16, No. 19, 3879

\bibitem[2000]{dhaw:00}
Dhawan, V., Mirabel, I.F., \& Rodriguez, L.F. 2000, ApJ, 543, 373

\bibitem[1996]{falh:96}
Falcke, H. 1996, ApJ, 464, L67

\bibitem[1996]{fal:96}
Falcke, H. and Biermann, P. L. 1996, A\&A, 308, 321

\bibitem[1999]{fal:99}
Falcke, H. and Biermann, P. L. 1999, A\&A, 342, 49 

\bibitem[1999]{fend:99}
Fender, R. P., Garrington, S. T., McKay, D. J. et al. 1999, MNRAS, 304, 865

\bibitem[1999]{gier:99}
Gierlinski, M., Zdziarski, A. A., Poutanen, J. et al. 1999, MNRAS, 309, 496

\bibitem[1988]{hjel:88}
Hjellming, R. M. and Johnston, K. J. 1988, ApJ, 328, 600

\bibitem[1996]{jaho:96}
Jahoda, K. et al. 1996, SPIE, 2808, 59

\bibitem[2000]{kota:00}
Kotani, T., Ebisawa, K., Dotani, T., Inoue, H., et al. 2000, ApJ, 539, 413

\bibitem[2001]{lee:01}
Lee, J. C., Schulz, N. S., Reynolds, C. S., Fabian, A. C. \& Blackman, E. G.
{\it Contributed talk at "X-ray Astronomy 2000", Palermo, Italy, Sept. 2000.} 
astro-ph/0012111

\bibitem[1994]{long:94}
Longair, M. S. 1994, High Energy Astrophysics, Vol 2, 2nd edition, CUP 

\bibitem[2001]{markf:01}
Markoff, S., Falcke, H. \& Fender. R. 2000, Accepted in  A\&AL, 
astro-ph/0010560

\bibitem[1999]{markw:99}
Markwardt, C. B., Swank, J. H., \& Taam, R. E. 1999, ApJ, 513, L37

\bibitem[1994]{mira:94}
Mirabel, I. F. \& Rodriguez, L. F. 1994, Nat, 371, 46

\bibitem[1999]{mira:99}
Mirabel, I. F. \& Rodriguez, L. F. 1999,  ARA\&A, 37, 409
 
\bibitem[1997]{morg:97}
Morgan, E. H., Remillard, R. A., \& Greiner, J. 1997, ApJ, 482, 993

\bibitem[1999]{muno:99}
Muno, M. P., Morgan, E.H., \&  Remillard, R. A. 1999, ApJ, 527, 321

\bibitem[2000]{naik:00}
Naik, S. \& Rao, A.R. 2000, A\&A, 362, 691 

\bibitem[1998]{nar:98}
Narayan, R., Mahadevan, R. \& Quataert, E. 1998, {\em The Theory of 
Black Hole Accretion Disks}, CUP

\bibitem[2000]{rao:00}
Rao, A. R., Naik, S., Vadawale, S. V., \& Chakrabarti, S. K. 2000, A\&A,
 360, L25

\bibitem[2000]{ryp:00}
Rao, A. R., Yadav, J. S. \& Paul, B. 2000, ApJ, 544, 443

\bibitem[2000]{rv:00}
Rao, A. R. and  Vadawale, S. V. 2000, Presented at the 4th Integral
Workshop, Alicante, Sept. 4-8, 2000.

\bibitem[1982]{rey:82}
Reynolds, S. P. 1982, ApJ, 256, 13

\bibitem[1998]{roth:98}
Rothschild, R.E. et al. 1998, ApJ, 496, 538

\bibitem[1980]{sun:80}
Sunyaev, E.A. \& Titarchuk, L. 1980, A\&A, 86, 121

\bibitem[1999]{trud:99}
Trudolyubov, S., Churazov, E., \& Gilfanov, M. 1999,  Ast. L., 25, 718

\bibitem[1999]{yadav:99}
Yadav, J. S., Rao, A. R., Agrawal, P. C. et al. 1999, ApJ, 517, 935

\end{thebibliography}
\end{document}